\documentstyle[times,pramana,epsf,floats]{ias}

\begin{document}
\mark{{Data Acquisition in the EUDET Project}{J. Mnich and M. Wing}}
\title{Data Acquisition in the EUDET Project}

\author{J. Mnich$^a$ and M. Wing$^b$}
\address{$^a$ Deutsches Elektronen-Synchrotron DESY, 
         Notkestr. 85, 22607 Hamburg, Germany \\
        $^b$ Dept. of Physics and Astronomy, University College London, Gower Street, London WC1E 6BT, UK
}
\pacs{2.0}
\abstract{
The goal of the EUDET project is the development and construction of infrastructure 
to permit detector R\&D for the International Linear Collider (ILC) with larger scale
prototypes. It encompasses major detector components: the vertex detector, the tracker 
and the calorimeters. We describe here the status and plans of the project with emphasis 
on issues related to data acquisition for future test beam experiments.
}

\maketitle

\section{Introduction}
EUDET~\cite{eudetweb} is a project supported by the European Union in the 
Sixth Framework Programme structuring the European Research Area~\cite{CORDIS}. 
The project is an Integrated Infrastructure Initiative (I3) which aims to create 
a coordinated European effort towards research and development for ILC detectors. 
The emphasis of the project is the creation and improvement of infrastructures to 
enable R\&D on detector technologies with larger prototypes. After establishing 
several new technologies to match the required ILC detector performances, the 
construction of and experimentation with larger scale prototypes to demonstrate the 
feasibility of these detector concepts is the next important step towards the design 
of an ILC detector. Such larger detectors generally require cooperation between 
several institutes and EUDET is intended to provide a framework for European and also 
Global collaboration.

The project comprises 31 European partner institutes from 12 different countries 
working in the field of High Energy Physics. In addition, 23 associated institutes 
will contribute to and exploit the EUDET research infrastructures. The project 
started in January 2006 and will run for four years providing additional funding of 
7\,MEuros from the European Union. In addition significant resources are committed 
by the participating institutes.

EUDET contributes to the development of larger prototypes of all detector components 
for which major R\&D efforts are ongoing: vertex and tracking detectors as well as 
calorimeters. The project is organised in three Joint Research Activities:
test beam infrastructure, infrastructure for tracking detectors and infrastructure for 
calorimeters, which are subdivided into several tasks addressing different detector types 
and technologies. The project is complemented by networking activities, the tasks of which 
include support for information exchange and common analysis tools as well as a transnational 
access scheme through which the use of the DESY test beam and, at a later stage, the 
exploitation of the infrastructures by European research groups is subsidised. 

With the increasing size and complexity of detector prototypes data acquisition issues 
become more and more important. For some of the EUDET infrastructures the development of a 
DAQ system is part of the project and first conceptual ideas have been developed. Even 
though it is certainly to early to design the final DAQ system it is instrumental to 
exchange ideas, homogenise concepts across sub-detector boundaries and thus prepare the 
ground for an integrated concept for the ILC detector. In EUDET a coherent DAQ approach is 
discussed for the large prototypes involved to facilitate combined test beam experiments.
Even though it was not part of the original project, discussions have started to evaluate
the feasibility in light of the very different demands of the detectors. It should also be 
noted that efforts on coherent DAQ schemes are very welcome to advance the concept of the 
Global Detector Network~\cite{GDN}.

\section{The Joint Research Activities}

\subsection{Test Beam Infrastructure}

This activity aims at improving the current test beam installation with a large bore magnet 
of up to about 1 Tesla and a low mass coil. The magnet, called PCMAG, is provided by KEK, 
one of the associated institutes. In addition a high resolution beam telescope made of pixel 
detectors using Monolithic Active Pixel Sensors (MAPS) is under development~\cite{Tobias}. 
Initially both 
devices will be constructed and used at the DESY test beam facility but they are 
transportable, as all EUDET infrastructures, and could be used later at other laboratories.

\begin{figure}[htbp]
\epsfxsize=8cm
\centerline{\epsfbox{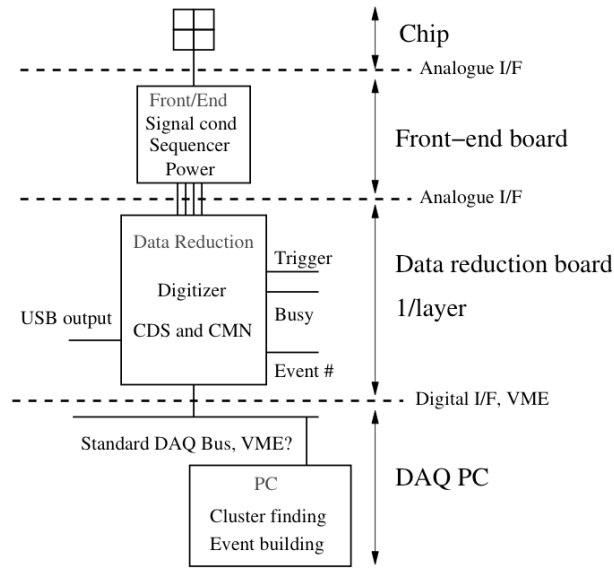}}
\caption{Design of the DAQ system for the high resolution pixel telescope.}
\label{fig:JRA1sketch}
\end{figure}

An important part of this project is the development of a DAQ system for the pixel
telescope. Fig.~\ref{fig:JRA1sketch} shows a first design of it. This task includes the 
design of front-end electronics and data reduction boards. It will be complemented by a 
special trigger logic unit. Some parts of the design, like the connection to the readout 
computers, are not yet decided. A first demonstrator set-up of the telescope is scheduled 
to become operational by mid 2007 and the full telescope by the end of 2008.

\subsection{Infrastructure for Tracking Detectors}

Both options for the ILC central tracking detector, a high resolution TPC and a large 
low-mass tracker consisting of silicon strip detectors, are addressed in this activity. The 
TPC activity centres around the development and construction of a large field cage, to be 
used inside PCMAG to test various options of micro pattern gas detectors which are under 
study for the gas amplification at the end-plates. For the silicon tracking option  
studies will concentrate on the design of a large and light mechanical structure, the cooling 
aspects as well as on front-end electronics development.

Together with the TPC field cage a general purpose readout system to be used with different 
end-plate technologies will be provided. The design of this readout is based on existing 
technologies, namely the ALTRO chip developed for the
ALICE experiment~\cite{altro}, which can provide the required high number 
of channels at low cost.
It has also the potential to be further developed and tailored to the requirements of the
ILC TPC using new high integration technologies as they become available.
The TPC readout system will be complemented by an adequate DAQ system.
This infrastructure is schedule to be ready for first test beam measurements beginning of 
2008.

\subsection{Infrastructure for Calorimeters}

This part of EUDET comprises the construction of a fully equipped module of the 
electromagnetic calorimeter, a versatile stack for testing technologies for the hadron 
calorimeter as well as calibration and sensor test devices for the forward calorimeter. 
The development of front-end electronics and a DAQ system for the calorimeters also belong 
to the project.

\begin{figure}[htbp]
\epsfxsize=8cm
\centerline{\epsfbox{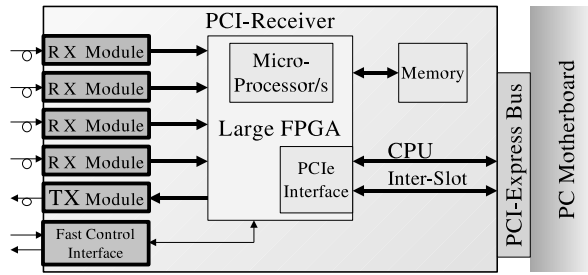}}
\caption{Off-detector receiver design for calorimeters.}
\label{fig:JRA3sketch}
\end{figure}

A conceptual design of the a DAQ system to be used with the electromagnetic and hadron 
calorimeters exists. It is flexible and can be adapted to different options of the readout 
electronics. Commercial products are used to ensure the system is inexpensive, scalable and 
maintainable. Fig.~\ref{fig:JRA3sketch} shows the concept of the off-detector receiver card. 
These cards will be mounted directly on PCI buses of the DAQ computers. This concept is 
expected to provide a high-speed generic DAQ card available in 2009 for test beam experiments.

\section{Conclusions}

Within the EUDET project, infrastructures for coming ILC detector R\&D with larger 
prototypes will be developed and constructed in the next years. DAQ systems for the 
calorimeters, vertex and tracking detectors are part of the project, which will permit 
detailed test beam experiments in a few years. Efforts have started to investigate if the 
concepts for these DAQ systems can be homogenised despite the partially diverging requirements 
on the R\&D issues to be addressed by the different detectors. Obviously, any modification 
and enlargement of the DAQ systems planned has to be accommodated within the time frame and 
the resources of EUDET. The advantages and possible benefits are, however, numerous ranging 
from combined test beam experiments to the valuable experience to be gained for the ILC 
detector.

 \section*{Acknowledgments}

This work is supported by the Commission of the European Communities under the 6th Framework Programme 
'Structuring the European Research Area', contract number RII3-026126.

\end{document}